\documentclass[paper,11pt]{JHEP}
\usepackage[centertags]{amsmath}
\usepackage{amsfonts} \usepackage{amssymb} \usepackage{amsthm}
\allowdisplaybreaks[1]
\usepackage{graphicx}
\newcommand{\Z}{{\mathbb Z}}

\newcommand{\eq}[1]{Eq.~(\ref{#1})}

\newcommand{\ii}{\mathrm{i}}
\newcommand{\ee}{\mathrm{e}}

\newcommand{\vev}[1]{\langle #1 \rangle}
\newcommand{\Tr}{\mathrm{Tr}\,}

\newcommand{\diag}{\mathrm{diag}\,}

\newcommand{\cE}{{\mathcal{E}}}

\newcommand{\cF}{{\mathcal{F}}}

\newcommand{\cN}{{\mathcal{N}}}
\newcommand{\cO}{{\mathcal{O}}}

\newcommand{\one}{{\rm 1\kern -.9mm l}}

\newcommand{\Pf}{\mathrm{Pf}}

\newcommand{\ft}[2]{{\textstyle\frac{#1}{#2}}}
\newcommand{\spin}[2]{\big[^{#1}_{#2}\big]}

\title{The exact 8d chiral ring from 4d recursion relations
}

\author{M. Bill\`o$^1$, M. Frau$^{1}$, L. Gallot$^{2}$, A. Lerda$^{3}$
\\
\vskip 0.2cm
$^1$ Dipartimento di Fisica Teorica, Universit\`a di Torino\\
and I.N.F.N. - sezione di Torino \\
Via P. Giuria 1, I-10125 Torino, Italy\\
\vskip 0.2cm
$^2$LAPTH, Universit\'e de Savoie, CNRS\\
9, Chemin de Bellevue,
74941 Annecy le Vieux Cedex, France\\
\vskip 0.2cm
$^3$Dipartimento di Scienze e Tecnologie Avanzate, Universit\`a del Piemonte
Orientale\\
and I.N.F.N. - Gruppo Collegato di Alessandria - sezione di Torino\\
Viale T. Michel  11, I-15121 Alessandria, Italy\\
\vspace{0.25cm}
\email{billo,frau,lerda@to.infn.it; laurent.gallot@lapp.in2p3.fr} 
}
\abstract{We consider the local F-theory set-up corresponding to four D7
          branes in type I$^\prime$ theory, in which the exact axio-dilaton background $\tau(z)$ is
          identified with the low-energy effective coupling of the four-dimensional
          $\cN=2$ super Yang-Mills theory with gauge group SU(2) and $N_f=4$ flavours
          living on a probe D3 brane placed at position $z$. Recently, an
          intriguing relation has been found between the correlators forming the
          chiral ring of the eight-dimensional theory on the D7 branes and 
          the large-$z$ expansion of the $\tau$ profile. 
          Here we apply to the SU(2) $N_f=4$ theory some recursion techniques
          that allow to derive the
          coefficients of the large-$z$ expansion of $\tau$ in terms of modular functions
          of the UV coupling $\tau_0$. In this way we obtain exact expressions
          for the elements of the eight-dimensional chiral ring that resum their instanton
          expansions, previously known only up to the first few orders by means of localization
          techniques.}

\preprint{DFTT/16/2011\\LAPTH 024/11}

\begin{document}

\section{Introduction}
\label{sec:intro}
F-theory is a very interesting framework for building string models that may be potentially
relevant for phenomenology (for reviews see, for instance, Ref.s~\cite{Donagi:2008ca,Heckman:2010bq}). 
It also represents an intriguing arena from the more
formal point of view, as it is supposed to incorporate the non-perturbative
corrections of type IIB string compactifications by geometrizing them in a very non-trivial
way.

By considering local type I$^\prime$ models containing D7 branes and an O7 plane, a
remarkable relation has been 
recently pointed out \cite{Billo:2010mg,Billo':2011uc,Fucito:2011kb} between the
profile of the axio-dilaton field $\tau$ in F-theory and certain correlators in
the eight-dimensional gauge theory living on the D7 branes that provide its microscopic
description. This relation reads 
\begin{equation}
\label{magic1}
\tau(a) = \tau_0 + \frac{1}{2\pi\ii} \Big\langle\log\det\left(1 - \frac ma\right)\Big\rangle
= \tau_0 -  \frac{1}{2\pi\ii} \sum_{\ell=1}^\infty \frac{1}{2\ell} \frac{\vev{\,\Tr
m^{2\ell}\,}}{a^{2\ell}}~. 
\end{equation} 
Here $a= z/(2\pi\alpha')$, where $z$ is the complex coordinate transverse to
the D7 world-volume. Moreover, $m$ is a complex scalar field
belonging to the 8$d$ chiral supermultiplet
that contains the massless degrees of freedom of the open strings attached to the D7 branes.
This multiplet transforms in the adjoint representation of $\mathrm{SO}(2 N_f)$ if there are
$N_f$ D7 branes, so it is an antisymmetric matrix and the traces of odd powers
in the expansion of the logarithm above vanish. The vacuum expectation values
in \eq{magic1} are taken with respect to the D7 brane world-volume theory.
Finally, we remark that $\tau$ depends on the vacuum expectation values
$m_i$ of the adjoint field $m$: 
\begin{equation}
\label{midef}
\vev{\, m\,} = \frac{1}{\sqrt{2}}\,\diag \big(m_1, m_2, \ldots, -m_1, -m_2,\ldots\big)~.
\end{equation}
The parameters $m_i = z_i/(2\pi\alpha')$ correspond to the locations $z_i$ of
the D7 branes when they are displaced from the orientifold plane.

In this paper we will focus on the case in which there are $N_f=4$ D7 branes, supporting an
$\mathrm{SO}(8)$ gauge theory in eight dimensions. This is the local limit of F-theory 
considered long ago by Sen in Ref.~\cite{Sen:1996vd}, where he proposed that the exact profile 
of $\tau$ is given by the effective low-energy coupling of the
$\mathcal{N}=2$ super Yang-Mills theory in four dimensions 
with gauge group SU(2), $N_f=4$ fundamental flavours and $\tau_0$ as UV coupling. 
This non-trivial relation can be understood \cite{Banks:1996nj} by considering a D3
brane probe of the geometry created by the D7 branes and the O7 plane: indeed, the D3 branes 
supports an $\mathrm{Sp}(1)\sim\mathrm{SU}(2)$ gauge theory with four flavours (plus a
decoupled hypermultiplet in the antisymmetric representation), 
and its gauge kinetic term couples to the axio-dilaton field. 
{From} this perspective, the parameters $m_i$ are the flavour
masses. 
When the complex scalar field $\phi$ that belongs to the gauge multiplet
on the D3 takes a vacuum expectation value 
\begin{equation}
\label{avev}
\vev{\,\phi\,} = (a,-a)~,
\end{equation}
{\it{i.e.}} when the D3 brane and its orientifold image are placed 
at $z= \pm 2\pi\alpha'a$, the exact axio-dilaton 
$\tau(a)$ represents the effective coupling of this gauge theory. As such, it is
encoded in the corresponding Seiberg-Witten (SW) curve \cite{Seiberg:1994rs,Seiberg:1994aj} of
which it describes the complex structure parameter.

\eq{magic1} was put forward in Ref.~\cite{Billo:2010mg} based on the computation of
the first few D-instanton corrections to the D3 coupling in the D3/D7 system
as a series in the non-perturbative parameter $q=\ee^{\ii\pi\tau_0}$. These
corrections were found to match the first few terms in the $q$-expansion of the 8$d$ chiral
ring elements obtained in Ref.~\cite{Fucito:2009rs} via localization techniques. In
\cite{Billo':2011uc} the same relation has been understood entirely from the
D7 brane point of view by showing how the D-instantons that correct the chiral
ring correlators also modify the source terms for $\tau$, hence its profile.
Adopting this point of view, in Ref.~\cite{Fucito:2011kb} the relation has been proven at
all instanton orders and extended to any number of D7 branes in presence of an
O7 plane, both in a flat background and in a $\mathbb{R}^4/\mathbb{Z}_2$ orbifold. 

The exact expression of $\tau$ encoded in the SW curve \emph{implicitly}
contains, via \eq{magic1}, all information about the 8$d$ chiral ring
correlators. However, to extract the exact expression of a given correlator, we
must be able to single out a specific term in the $1/a$ expansion of $\tau$.
This is not trivial, but it can be done systematically by using recursive
techniques, akin to the Matone relation \cite{Matone:1995rx}. Here we will
rediscuss this type of recursions, gathering an understanding that allows us to
apply them also to the $N_f=4$ case, where the structure of the SW curve is
complicated by the presence of several different invariants constructed with the
flavour masses. In this way we are able to obtain exact expressions for the 8$d$
chiral ring elements that resum their instanton expansions, previously known
only up to the first few orders. We find this a remarkable by-product of the
deep relation between the 4$d$ effective physics on the D3 brane and the 8$d$
theory on the ``flavour'' D7 branes encoded in \eq{magic1}. Let us note that
for conformal theories there exists also another recursive approach, based on the modular anomaly
equation \cite{Minahan:1997ct,Bershadsky:1993cx}, which allows to partially fix
the coefficients of the large-$a$ expansion of the effective coupling $\tau$;
this approach, which has been used in Ref.~\cite{Minahan:1997if} for the
so-called $\cN=2^*$ theory (also known as mass deformed $\cN=4$ theory) where
there is a single mass invariant, could also be applied to the $N_f=4$ model,
but it is less efficient than the one we are going to discuss.

The structure of this paper is as follows: in Section \ref{recrel} we 
discuss in general the recursion relations for rank one $\cN=2$ theories, and describe the
procedure to follow when the SW curve is not in a factorized form. In Section \ref{recnf4}
we discuss the SU(2) $N_f=4$ theory and show how to obtain from its SW curve a recursion relation
that yields the various correlators of the ``flavour'' theory, whose properties 
are presented in Section
\ref{sec:results} together with comments and concluding remarks. Finally, in the Appendices
we give some more technical details and discuss the recursion relation in 
the $\cN=2^*$ model seen as a particular case of the $N_f=4$ theory.

\section{Recursion relations for rank one $\cN=2$ theories}
\label{recrel}
The SW curve for ${\cal N}=2$ super Yang-Mills theories with gauge group 
$\mathrm{SU(2)}$ is a torus and can be thus described as an algebraic surface 
via a cubic equation of the form
\begin{equation}
\label{genfact}
y^2= \big(x-{\cal E}_1(z)\big) \big(x-{\cal E}_2(z)\big) \big(x-{\cal E}_3(z)\big)~,
\end{equation}
where the precise expression for the  roots ${\cal E}_{\ell}$ depends on the 
gauge-invariant coordinate on the Coulomb moduli space
\begin{equation}
\label{defu}
u = \vev{\,\Tr\phi^2\,}~,
\end{equation}
on the masses (if there is matter) and on the dynamically
generated scale $\Lambda$ (or the bare UV coupling $\tau_0$ in the conformal cases). All these
dependencies are here summarized by the variable $z$. The complex structure parameter
$\tau$ of the torus describes the complexified IR coupling of the gauge theory according to
\begin{equation}
\label{tauYM}
 \tau=\frac{\theta_{\mathrm{eff}}}{\pi}+\frac{8\pi\ii}{g^2_{\mathrm{eff}}}
\end{equation}
where $\theta_{\mathrm{eff}}$ and $g_{\mathrm{eff}}$ are the
$\theta$-angle and the Yang-Mills coupling constant at low-energy, and is related 
to the anharmonic ratio $\kappa$ of the roots as follows
\begin{equation}
\label{kappais}
\kappa=\frac{\cE_3(z) - \cE_2(z)}{\cE_1(z) - \cE_2(z)} = \frac{\theta_2^4(\tau)}{\theta_3^4(\tau)}~,
\end{equation}
where the $\theta$'s are the Jacobi $\theta$-functions (see Appendix \ref{app:formulae} for
our conventions).
In the semiclassical regime, {\it i.e.} when $u$ is large, we have
\begin{equation}
 \label{ua}
  u \sim \Tr \vev{\phi}^2 = 2 a^2~,
\end{equation}
where the second equality follows from \eq{avev}. At a generic point
$u$ on the moduli space we have a symplectic section $(a(u),a_D(u))$ given by
the periods of the SW differential, such that
\begin{equation}
\label{aad}
\frac{\partial a}{\partial u} = \omega_1~,~~~
\frac{\partial a_D}{\partial u} = \omega_2~,
\end{equation}
where $\omega_{1}$ and $\omega_2$ are the periods of the torus with 
\begin{equation}
\label{tautoaad}
\tau = \frac{\omega_2}{\omega_1} = \frac{\partial a_D}{\partial a}~.
\end{equation}
The low-energy physics can be described by an effective theory for an abelian 
multiplet with lowest component $a$ and a prepotential $\cF(a)$ such that
\begin{equation}
\label{preptoad}
a_D = \frac{1}{2\pi\ii}\frac{\partial \cF(a)}{\partial a}~,
\end{equation}
which implies that 
\begin{equation}
\label{tautoF}
\tau = \frac{1}{2\pi\ii} \frac{\partial^2 \cF(a)}{\partial a^2}~.
\end{equation}

The SW curve (\ref{genfact}) encodes the exact expressions for the physical
quantities in the corresponding gauge theory, including its effective coupling
$\tau$. These quantities admit a large-$a$ expansion which exhibits, beside the
tree-level and perturbative terms, also non-perturbative contributions from all
instanton sectors which can be computed directly using localization techniques
in the multi-instanton calculus \cite{Nekrasov:2002qd}. To extract the instanton
expansion from the SW curve one has to determine $\tau(a)$. This can be done by
first obtaining the expression of $\tau(u)$ by inverting \eq{kappais}, and then
by determining $u$ as a function of $a$ by inverting the first relation in
\eq{aad}. This procedure is straightforward but may become rather cumbersome in
practice.

A more efficient way to proceed is to write the prepotential $\cF(a)$, and hence
$\tau(a)$, as an expansion for large $a$ with unknown coefficients and to obtain
a recursion relation for the latter. This can be done by expanding the right
hand side of \eq{kappais} around a specific value of $\tau$ corresponding to the
semiclassical limit, and the left hand side of \eq{kappais} around particular
values of the roots $\cE_\ell$ of the SW curve that correspond to this limit.
This is basically the idea behind the recursion relation originally devised in
Ref.s~\cite{Matone:1995rx,Bonelli:1996qc} for the pure $\mathrm{SU}(2)$ theory.
In this case the SW curve is \cite{Seiberg:1994rs}
\begin{equation}
 y^2=\big(x-\hat u\big) \big(x-\hat\Lambda^2\big) \big(x+\hat\Lambda^2\big)~,
\label{puresu2}
\end{equation}
so that the anharmonic ratio of the roots defined in \eq{kappais} is
\begin{equation}
 \hat\kappa=\frac{-2\hat\Lambda^2}{\hat u-\hat\Lambda^2}~.
\label{kappasu2}
\end{equation}
In these expressions we have introduced a ``hat'' sign to take into account the fact that
the complex structure of the SW curve (\ref{puresu2}) turns out to be 
related to the gauge theory parameters by \cite{Seiberg:1994rs}
\begin{equation}
\label{tauYM2}
 \hat\tau=\frac{\theta_{\mathrm{eff}}}{2\pi}+\frac{4\pi\ii}{g^2_{\mathrm{eff}}}~,
\end{equation}
as opposed to \eq{tauYM}; moreover, the parameter $\hat u$ is
related to the vacuum expectation value $a$ in the semi-classical regime by $\hat u \sim a^2/2$,
to be contrasted with \eq{ua}%
\footnote{\label{foot_norm}
To avoid this change of conventions and normalizations, 
in place of the curve (\ref{puresu2}) one could use for the pure SU(2) theory the isogenic SW curve
\begin{equation*}
 y^2=\big(x-u+\sqrt{u^2-\Lambda^4}\big)\,x\,\big(x-u-\sqrt{u^2-\Lambda^4}\big)
\end{equation*}
with complex structure $\tau=2\hat\tau$ in agreement with \eq{tauYM}. 
The anharmonic ratio of the roots would then read 
\begin{equation*}
\label{altkappa}
\kappa = \frac{u - \sqrt{u^2 - \Lambda^4}}{u + \sqrt{u^2 - \Lambda^4}}~.
\end{equation*}}.

In Ref.~\cite{Matone:1995rx} it was shown that that the quantity
$\hat u/\hat\Lambda^2$ satisfies a differential equation
whose solution is
\begin{equation}
\label{ulambda}
 \frac{\hat u}{\hat\Lambda^2}= 1-2\,\frac{\theta_3^4(\hat\tau)}{\theta_2^4(\hat\tau)}~;
\end{equation}
from this result the identification $\hat\kappa={\theta_2^4(\hat\tau)}/{\theta_3^4(\hat\tau)}$ 
immediately follows.

One extra ingredient that is needed to proceed is the relation between $u$ (or
$\hat u$) and $a$, which generalizes the classical one given in \eq{ua}. Such a
relation is provided \cite{Matone:1995rx} through the prepotential $\cF(a)$ by
means of
\begin{equation}
 u(a) = 2\Lambda \frac{\partial\cF(a)}{\partial\Lambda}~.
\label{uLF}
\end{equation}
Inserting this in the left hand side of \eq{ulambda} and using \eq{tautoF} in
the right hand side, one obtains a non-trivial equation for the prepotential
from which, by expanding in inverse powers of $a$, one can derive a recursion
relation for the coefficients of this expansion. For dimensional reasons these
coefficients correspond to different powers of $\Lambda^4$, {\it{i.e.}} to
different instantonic sectors%
\footnote{Recall that corrections from the sector with instanton number $k$ are weighted
by $\Lambda^{b_1 k}$, where $b_1$ is the 1-loop coefficient of the $\beta$-function. 
For the pure SU(2) theory we have $b_1=4$.}, 
and hence this recursion relation allows to reconstruct the higher instanton contributions 
starting from the lower ones. 

Things are a bit different in conformal theories. In this case the relation
(\ref{uLF}) is replaced by
\begin{equation}
 u(a) = 2 q \frac{\partial\cF(a)}{\partial q}
\label{utoF}
\end{equation}
where 
\begin{equation}
\label{defq}
 q=\ee^{\pi\ii\tau_0}
\end{equation}
with $\tau_0$ being the bare UV coupling. Inserting \eq{utoF} in the left hand
side of \eq{kappais} and replacing in the right hand side $\tau$ via
\eq{tautoF}, one generates a recursion relation for the coefficients of the
large-$a$ expansion of the prepotential in which the $q$ dependence is
\emph{exact}. 

However, not always the SW curve is given or known in the factorized form
(\ref{genfact}) considered so far. For example, for the SU(2) theories with
$N_f\leq4$ massive flavours the SW curves are written as cubic polynomials in a
non-factorized form \cite{Seiberg:1994aj} for which it is not easy or practical%
\footnote{Even if, in principle, it is always possible via the Cardano formula.} 
to find the three roots $\cE_\ell$. Thus, in these cases the recursion relation
cannot be directly obtained by applying the above procedure. Nevertheless, a simple
generalization exists and a recursion relation can be implemented also in these
cases%
\footnote{The technique we describe here is similar to the one used in the first
part of Ref.~\cite{Minahan:1997ct} to find the instanton expansion of toroidally 
compactified non-critical strings.}. In fact, by shifting the variable $x$ if needed, it is always possible to
put a cubic polynomial in a Weierstra\ss~form:
\begin{equation} 
\label{genwei}
y^2 = x^3 - \frac{G_2(z)}{4}\, x - \frac{G_3(z)}{4}~.
\end{equation} 
In this description the complex structure $\tau$ can be directly related 
to the coefficients $G_2$ and $G_3$
by forming the combination
\begin{equation}
 \label{Jdef}
 J = \frac{G_2^3(z)}{G_2^3(z) - 27\, G_3^2(z)}~,
 \end{equation}
and identifying it with the ``absolute modular invariant'' by writing
\begin{equation}
\label{Jtotau}
J = \frac{E_4^3(\tau)}{E_4^3(\tau) - E_6^2(\tau)} 
\end{equation}
where $E_4$ and $E_6$ are the Eisenstein series of weight 4 and 6, respectively.
By equating the right hand sides of Eq.s~(\ref{Jdef}) and (\ref{Jtotau}) we obtain the relation
between $\tau$ and $u$, and then we can proceed as described above and establish a recursion
relation by exploiting either \eq{utoF} or \eq{uLF} depending on whether or
not the theory is conformal.

In the next section we will apply this method to the SU(2) theory with $N_f=4$ massive flavours
whose SW curve is known in a non-factorized form \cite{Seiberg:1994aj}, and explicitly derive a large-$a$ expansion of its effective coupling $\tau$
in which each coefficient is determined \emph{exactly} as a function of $q$ by means of a recursion
relation. As explained
in the Introduction, via \eq{magic1} this is tantamount to finding the exact 
expression of the elements of the 8$d$ chiral ring on the ``flavour'' D7 branes.

\section{The SU(2) $N_f=4$ theory} 
\label{recnf4} 
The $\mathrm{SU}(2)$ $N_f=4$ theory for vanishing masses is conformal and the corresponding
SW curve is just a torus of complex structure $\tau_0$, representing the UV coupling
\cite{Seiberg:1994aj}. 
Such a torus can be given a simple description as the locus of a factorized 
cubic equation in Weierstra\ss~form
\begin{equation}
\label{cub0}
y^2 = x^3 - \frac{g_2}{4} x - \frac{g_3}{4} = (x - e_1)(x - e_2)(x - e_3)~,
\end{equation}
where the three roots $e_\ell$, satisfying $e_1 + e_2 + e_3 =0$, are the
following functions of $\tau_0$:
\begin{equation}
\label{eiare}
e_1 = \frac 13 \left(\theta_3^4 + \theta_4^4\right)~,~~~
e_2 = -\frac 13 \left(\theta_3^4 + \theta_2^4\right)~,~~~
e_3 = -\frac 13 \left(\theta_4^4 - \theta_2^4\right)~.
\end{equation}
with $\theta_a$ being the Jacobi $\theta$-functions%
\footnote{For brevity, here and in the following, 
when no modular variable is indicated and no
confusion is possible, we always understand that the modular functions are 
evaluated at $\tau_0$; for example $\theta_a\equiv\theta_a(\tau_0)$. We refer to
Appendix \ref{app:formulae} for our conventions and useful relations.}.
The coefficients $g_2$ and $g_3$ can then be expressed in terms of the
Eisenstein series $E_i$ as
\begin{equation}
\label{g23are}
g_2 = \frac 43 \,E_4~,~~~
g_3 = \frac{8}{27}\, E_6~,
\end{equation}
so that the absolute modular invariant $J_0\equiv J(\tau_0)$, 
in accordance with Eq.s~(\ref{Jdef}) and (\ref{Jtotau}), becomes 
\begin{equation}
\label{j0is}
\frac{1}{J_0} = 1 - 27\, \frac{g_3^2}{g_2^3} = 1 - \frac{E_6^2}{E_4^3}~.
\end{equation}
Notice that all the functions of $\tau_0$ involved in the above definitions are expressible 
as power series in $q$ defined in \eq{defq}.

When masses are turned on, the equation of the curve is modified as described in
Ref.~\cite{Seiberg:1994aj}. The result is still a cubic polynomial which can be written as
\begin{equation}
\label{curvemassive}
 y^2 =  W_1 W_2 W_3 + A \Big[W_1 T_1 (e_2 - e_3) + W_2 T_2 (e_3 - e_1) 
+ W_3 T_3 (e_1 - e_2)\Big]  - A^2 N~,
\end{equation}
where%
\footnote{Note that $A^2$ is proportional to the discriminant of the cubic 
equation and can be written as
\begin{equation*}
\label{A2is}
A^2 = \frac{1}{16} (g_2^3 - 27 g_3^2) = \frac{4}{27}(E_4^3 - E_6^2)~.
\end{equation*}}
\begin{equation}
 \label{Adef}
 A = (e_1 - e_2)(e_2 - e_3)(e_3 - e_1) = 16\, \eta^{12}~,
\end{equation}
with $\eta$ being the Dedekind $\eta$-function and, for $\ell=1,2,3$, 
\begin{equation}
 \label{swW}
  W_\ell = x - e_\ell\, \tilde u - e_\ell^2\, R~,
\end{equation}
with
\begin{equation}
 \label{utdef}
  \tilde u = u - \frac{e_1}{2}\,R~.
\end{equation}
Here $R$, $T_\ell$ and $N$ are invariants of the flavour group SO$(8)$
that are, respectively, quadratic, quartic and sextic in the masses $m_i$:
\begin{equation}
 \label{invariants}
  \begin{aligned}
     R&= \frac{1}{2}\,\sum_i m_i^2~,\\
     T_1&=\frac{1}{12}\,\sum_{i<j}m_i^2m_j^2-\frac{1}{24}\,\sum_im_i^4~,\\
     T_2&=-\frac{1}{24}\,\sum_{i<j}m_i^2m_j^2+\frac{1}{48}\,\sum_im_i^4-\frac{1}{2}\,\mathrm{Pf}m~,\\ 
     N&=\frac{3}{16}\sum_{i<j<k}m_i^2m_j^2m_k^2-\frac{1}{96}
\,\sum_{i\not=j}m_i^4m_j^2+\frac{1}{96}\,\sum_i m_i^6 
 \end{aligned}
\end{equation}
with $\mathrm{Pf}m=m_1m_2m_3m_4$.
The third quartic invariant $T_3$ is not independent, rather it is defined  
through the relation $T_1 + T_2 + T_3 =0$.  

When the masses are set to zero, it is straightforward to see that 
\eq{curvemassive} reduces to the equation%
\footnote{With respect to \eq{cub0}, the roots are rescaled by $u$; 
this does not affect the complex structure and the absolute modular invariant.} 
of a torus of complex parameter $\tau_0$. 
For non-zero masses, by a suitable shift%
\footnote{Explicitly, $x~\to~x+\frac{2}{9}E_4 R$.} in $x$,
the curve can still be cast in the Weierstra\ss~form (\ref{genwei})
with coefficients $G_2$ and $G_3$ depending on $u$, on the flavour invariants, 
and on $q$. Their explicit expressions are
\begin{equation}
 \begin{aligned}
  G_2&=\frac{4}{3}\,E_4\,\tilde u^2+\frac{8}{9}\,E_6\,R\,\tilde u+\frac{4}{27}\,E_4^2\,R^2
+12\,A\big(e_1\,T_2-e_2\,T_1\big)~,\\
G_3&=\frac{8}{27}\,E_6\,\tilde u^3+\frac{8}{27}\,E_4^2\,R\,\tilde u^2
+ \frac{8}{81}\,E_4E_6\,R^2\tilde u-\frac{8}{729}\big(E_4^3-2E_6^2\big)R^3
+4\,A^2 N \\
&~~~+\frac{4}{3}\,AE_4\big(e_1\,T_2-e_2\,T_1\big)R
-\frac{4}{3}\,A\Big(E_4\big(T_1-T_2)+9\,e_3\big(e_1\,T_1-e_2\,T_2\big)\Big)\tilde u~.
 \end{aligned}
\label{g2g3}
\end{equation}
It is easy to check that in the massless limit we have
\begin{equation}
 G_2~\to~G_2^{(0)}=g_2\,u^2~,~~~G_3~\to~G_3^{(0)}=g_3\,u^3
\label{g230}
\end{equation}
with $g_2$ and $g_3$ given in \eq{g23are}.

The modular invariant $J$ can then be explicitly determined in terms of 
$u$, of the flavour invariants and of $q$ by \eq{Jdef}, which we rewrite as
\begin{equation}
\label{jtoG}
\frac{1}{J} = 1 - 27\, \frac{G_3^2}{G_2^3}~.
\end{equation}
The complex structure $\tau$, namely the exact
IR complexified gauge coupling, is in turn implicitly determined by
the modular invariant, to which it is related by \eq{Jtotau}, that we rewrite as
\begin{equation}
\label{jis}
\frac{1}{J} =  1 - \frac{E_6^2(\tau)}{E_4^3(\tau)}~.
\end{equation}
Comparing these two expressions for $J$ we can establish a relation between $u$ and $\tau$. However, in order to make contact with the standard field theory results, 
we have to write everything in terms of $a$. The 
ingredients that are needed for this purpose, namely the functions $u(a)$ and $\tau(a)$, 
are provided by the prepotential $\cF(a)$ through Eq.s~(\ref{utoF}) and (\ref{tautoF}), respectively.
At this point, when both Eq.s~(\ref{jtoG}) and (\ref{jis}) give $J$ as a function of $a$, 
we can proceed in two distinct ways. On the one hand, we can expand \eq{jis} in powers of
\begin{equation}
 \label{Qis}
Q=\ee^{\pi\ii\tau}~,
\end{equation}
and \eq{jtoG} in powers of $q$, and then compare the two expansions, thus finding order by order 
a relation between $Q$ and $q$ in which the dependence on $a$ (and the mass invariants)
is exact \cite{Billo:2010mg}. 
On the other hand, we can expand both Eq.s~(\ref{jtoG}) and (\ref{jis}) in (inverse) powers
of $a$ and, by comparing the two expansions, obtain a recursion relation for their coefficients
which allows to determine exactly the full dependence on $q$. 

We now show that using the first approach we can easily reconstruct the 1-loop
corrections to the gauge coupling constant and the prepotential of the $N_f=4$ theory. 
Later, we will exploit the second approach and study the recursion relation.

\subsection{Tree-level and 1-loop terms}
\label{subsec:tree1loop}
Expanding \eq{jis} for small $Q$, we obtain
\begin{equation}
 \frac{1}{J}= 1728\,Q^{2}+\cO(Q^{4})~;
\end{equation}
likewise, using \eq{g2g3} and then expanding \eq{jtoG} for small $q$, we get
\begin{equation}
 \frac{1}{J}=1728\,q^2\,\Big(1-\frac{2R}{u}+\frac{R^2+6T_1}{u^2}-
\frac{4N+2RT_1}{u^3}+\frac{(T_1+2T_2)^2}{u^4}\Big)+\cO(q^4)~.
\end{equation}
Equating these two expressions and using the classical approximation (\ref{ua}) to replace
$u$ with $2a^2$, we deduce that
\begin{equation}
\begin{aligned}
 Q&=q\,\sqrt{1-\frac{R}{a^2}+\frac{R^2+6T_1}{4a^4}-
\frac{4N+2RT_1}{8a^6}+\frac{(T_1+2T_2)^2}{16a^8}}+\cO(q^2)\phantom{\Bigg|}\\
&=q\,\sqrt{1-\frac{\sum_im_i^2}{a^2}+\frac{\sum_{i<j}m_i^2m_j^2}{4a^4}-
\frac{\sum_{i<j<k}m_i^2m_j^2m_k^2}{8a^6}+\frac{(\mathrm{Pf}m)^2}{16a^8}}+\cO(q^2)\phantom{\Bigg|}
\end{aligned}
\label{Qq}
\end{equation}
where the second line follows from the definitions (\ref{invariants}) of the mass invariants.
The square root in \eq{Qq} represents the complete 
1-loop correction to the UV coupling as a function of the
mass deformations and of the classical vacuum expectation value $a$. By taking the logarithm
of both sides and using \eq{midef}, after some simple algebra, we can rewrite \eq{Qq} as follows
\begin{equation}
\begin{aligned}
 \tau&
=\tau_0-\frac{1}{2\pi\ii}
\sum_{\ell=1}^\infty \frac{1}{2\ell} 
\frac{\Tr\vev{m}^{2\ell}}{a^{2\ell}}+\cO(q)~,
\end{aligned}
\label{tau1loop}
\end{equation}
which indeed is the correct expression for the gauge coupling constant of the massive
SU(2) $N_f=4$ theory in the 1-loop approximation. We can also write this result in terms
of the prepotential $\cF(a)$ 
in accordance with \eq{tautoF}. 
Introducing for later convenience
the quantities
\begin{equation}
 h_\ell^{(0)} = \frac{2^\ell}{(2\ell+1)(2\ell+2)}\,\Tr\vev{m}^{2\ell+2}~,
\label{hl0}
\end{equation}
for $\ell\geq0$, the prepotential that follows from \eq{tau1loop} is then
\begin{equation}
\label{fa0}
 \cF(a)=\pi\ii\tau_0\,a^2+\log\Big(\frac{a}{\Lambda}\Big) 
h_0^{(0)}-\sum_{\ell=1}^\infty\frac{1}{2\ell\,2^\ell}\,
\frac{h_\ell^{(0)}}{a^{2\ell}}+\cO(q)~,
\end{equation}
up to possible $a$-independent terms.
One can easily check that this agrees with the perturbative expression obtained with standard
field theory methods (see, for example, Ref.~\cite{D'Hoker:1999ft}) up to 
constant terms which can always be absorbed into a redefinition of the (arbitrary) scale $\Lambda$.

To obtain the $q$-dependent terms in the prepotential, one can go to higher order in the $q$-expansion, 
as discussed in Ref.~\cite{Billo:2010mg}. Alternatively, one can expand 
the modular invariant $J$ given by Eq.s~(\ref{jtoG}) and (\ref{jis}) 
in inverse powers of $a$ and by comparing the two expansions establish a recursion relation for their 
coefficients which fixes
the complete $q$-dependence. This is the approach we are going to discuss in the following.

\subsection{Initial condition}
\label{subsec:initial}
In order to successfully implement a recursion relation, we need to know,
as an initial condition, the exact expression in $q$ of the first sub-leading
term of $u(a)$ for large $a$. In other words, writing
\begin{equation}
\label{sublu}
 u(a) = 2 a^2 + 2 \lambda(q)\,R  + \cO(a^{-2})~,
\end{equation}
we need to determine the function $\lambda(q)$. To do this, we start by inverting the above
relation, obtaining
\begin{equation}
\label{subla}
a= \frac{u^{1/2}}{\sqrt{2}}-\frac{\lambda(q)\,R}{\sqrt{2}\,u^{1/2}}+\cO(u^{-3/2})~.
\end{equation}
{From} this it readily follows that the first period $\omega_1$ of the torus has the expansion
\begin{equation}
\label{subldadu}
\omega_1 = \frac{\partial a}{\partial u} =
\frac{1}{2\sqrt{2}\,u^{1/2}}+\frac{\lambda(q)\,R}{2\sqrt{2}\,u^{3/2}}+\cO(u^{-5/2})~.
\end{equation}
On the other hand, given the Weierstra\ss~form (\ref{genwei}), the period $\omega_1$ can be expressed as 
\cite{Masuda:1996xj}
\begin{equation}
\omega_1 = (48\, G_2)^{-{1}/{4}} ~F\big(\ft{1}{12},\ft{5}{12},1; \ft{1}{J}\big)~,
\label{omega1}
\end{equation}
where $F(a,b,c;z)$ is the hypergeometric function and the normalization has been chosen 
so that the leading behaviour $1/(2\sqrt{2}\,u^{1/2})$ is correctly reproduced.
{From} this expression we can obtain 
the term $\delta\omega_1$ linear in $R$ as a 
perturbation around the massless case; then, by writing it as
\begin{equation}
\label{deltaomega}
 \delta \omega_1 = \frac{\lambda(q)\,R}{2\sqrt{2}\,u^{3/2}}
\end{equation}
according to \eq{subldadu}, we can read off $\lambda(q)$. 
Let us now give some details. {From} \eq{g2g3} we easily find
\begin{equation}
 \begin{aligned}
\delta G_2 &\equiv G_2-G_2^{(0)}= G_2^{(0)}\Big(\frac{2E_6}{3E_4}-e_1\Big)\frac{R}{u}+\cdots~,\\
\delta G_3 &\equiv G_3-G_3^{(0)}= G_3^{(0)}\Big(\frac{E_4^2}{E_6}-\frac{3e_1}{2}\Big)\frac{R}{u}+\cdots~,
 \end{aligned}
\label{deltag2g3}
\end{equation}
where the dots stand for terms of higher order in the mass deformations which are not relevant for
our present purposes. Moreover, from \eq{jtoG} we get
\begin{equation}
 \delta\big(\frac{1}{J}\big)\equiv\frac{1}{J}-\frac{1}{J_0}=\frac{1-J_0}{J_0}\,
\Big(\frac{2\,\delta G_3}{G_3^{(0)}}
-\frac{3\,\delta G_2}{G_2^{(0)}}\Big)=\frac{2E_6\big(E_6^2-E_4^3\big)}{E_4^4}\,\frac{R}{u}+\cdots~.
\label{delta1J}
\end{equation}
On the other hand, by varying \eq{omega1} we obtain
\begin{equation}
 \delta\omega_1=\omega_1^{(0)}\Bigg[\!\!-\frac{\delta G_2}{4G_2^{(0)}}
+\partial_z\log F\big(\ft{1}{12},\ft{5}{12},1; z\big)\Big|_{z=\frac{1}{J_0}}\,\delta\big(\frac{1}{J}\big)
\Bigg]~.
\end{equation}
Taking into account that $F\big(\frac{1}{12},\frac{5}{12},1; \frac{1}{J_0}\big)=E_4^{1/4}$, and using 
Eq.s~(\ref{deltag2g3}) and (\ref{delta1J}), after some algebra involving the properties of the Eisenstein series
and the $\theta$-functions collected in Appendix \ref{app:formulae}, we get
\begin{equation}
 \delta\omega_1=\frac{R}{2\sqrt{2}\,u^{3/2}}\,\Big(\frac{e_1}{4}-\frac{E_2}{6}\Big)+\cdots
\label{omega10}
\end{equation}
Upon comparison with \eq{deltaomega}, we thus obtain
\begin{equation}
\label{lambdais}
\lambda(q) = \frac{e_1}{4} - \frac{E_2}{6} = \frac{1}{12} (\theta_3^4+\theta_4^4 -2 E_2) = 
-q\frac{\partial}{\partial_q}\log\big(\theta_3\theta_4\big)
~.
\end{equation}
Inserting this into \eq{sublu} and using the resulting expression for $u(a)$ in \eq{utoF}, 
we can easily derive the leading terms of the semiclassical expansion 
of the prepotential $\cF(a)$, namely
\begin{equation}
\label{expF0}
 \cF(a)= \pi\ii\tau_0\,a^2 - \log\big(\theta_3\theta_4\big) \,R+\cO(a^{-2})~.
\end{equation}
It is interesting to remark that this structure has also been obtained 
in Ref.~\cite{Marshakov:2009kj} using the AGT conjecture \cite{Alday:2009aq} and the Zamolodchikov formula for the 4-point conformal blocks in the two-dimensional Liouville theory%
\footnote{We observe that the result reported in Eq.~(33) 
of Ref.~\cite{Marshakov:2009kj} does not
respect the SO(8) flavour symmetry due the presence of the 
structure $(\sum_i m_i)^2$ which is not SO(8) invariant. However, as remarked in 
Ref.~\cite{Billo:2010mg}, this fact can be easily corrected by choosing a
different ``dressing factor'' in the AGT relation. With this choice the resulting 
SO(8) invariant expression for the
prepotential fully agrees with Eq.~(\ref{expF0}).}.

\subsection{Recursion relation}
\label{subsec:recursion}
By comparing the two expressions (\ref{fa0}) and (\ref{expF0}), we are immediately led to write
the following expansion for the complete prepotential
\begin{equation}
\label{expF}
 \cF(a)= \pi\ii\tau_0\,a^2 - \log\big(\theta_3\theta_4\big) \,R
+\log\Big(\frac{a}{\Lambda}\Big) h_0-\sum_{\ell=1}^\infty\frac{1}{2\ell\,2^\ell}\,\frac{h_\ell}{a^{2\ell}}~.
\end{equation}
The coefficients $h_\ell$, to be determined, are the generalizations of those 
defined in \eq{hl0} when non-perturbative instanton corrections are taken into account. 
Via \eq{tautoF}, the expansion (\ref{expF}) 
corresponds to writing the IR coupling $\tau$ as
\begin{equation}
 \label{tau}
\tau -\tau_0 = -\frac{1}{2\pi\ii}
\sum_{\ell=0}^\infty \frac{2\ell+1}{2^\ell}\,\frac{h_\ell}{a^{2\ell+2}}~.
\end{equation}
Comparing this expression with \eq{magic1}, we see that the coefficients 
$h_\ell$ are related to the elements of the  chiral ring of the eight-dimensional 
SO(8) theory by%
\footnote{We have written the expansion (\ref{expF}) in terms of the coefficient $h_\ell$ and
not of $\vev{\Tr m^{2l+2}}$ because the former turn out to be more 
convenient to exhibit the results in a compact way,
thanks to the fact that the ''modular anomaly'' equation they satisfy 
(to be discussed in the next section) takes a particularly simple form with this 
choice.}
\begin{equation}
\label{htochir}
h_\ell = \frac{2^\ell}{(2\ell+2)(2\ell+1)} \vev{\,\Tr m^{2\ell+2}\,}~.
\end{equation} 
The absence in the effective prepotential of any dependence on the scale $\Lambda$ other than that 
arising at 1-loop implies that
\begin{equation}
 h_0= h_0^{(0)}=R~,
\label{h00}
\end{equation}
or, equivalently, that 
$\vev{\,\Tr m^{2}\,}=\Tr\vev{m^{2}}$. This relation, which is explicitly confirmed by
instanton calculations \cite{Fucito:2009rs}, can be understood also as a simple consequence of the
scaling dimensions of the instanton moduli space which do not allow to generate any non-perturbative contribution to $\vev{\,\Tr m^{2}\,}$.

The next ingredient is obtained from \eq{utoF} which,
together with the expansion (\ref{expF}) and \eq{h00}, implies that 
\begin{equation}
\label{expu}
u(a) = 2 a^2 + 2 \lambda(q)\,h_0 - \sum_{\ell=1}^\infty\frac{1}{\ell\,2^\ell}\,\frac{q\partial_q h_\ell}{a^{2\ell}}
\end{equation}
with $\lambda(q)$ given in \eq{lambdais}.

Now, we are ready to establish the desired recursion relation. On the one hand, we expand $1/J$ around 
$\tau_0$, getting
\begin{equation}
\label{tayloriJ}
\delta\big(\frac{1}{J}\big)=\frac{1}{J}-\frac{1}{J_0}= \sum_{n=1}^\infty \frac{1}{n!} 
\left.\frac{\partial^n(1/J)}{\partial\tau^n}\right|_{\tau_0} (\tau - \tau_0)^n~.
\end{equation} 
The difference $(\tau - \tau_0)$ is parametrized as in \eq{tau}, while the
 derivatives of $1/J$ can be straightforwardly computed from \eq{jis} in terms
 of the derivatives of the Eisenstein series given in Appendix
 \ref{app:formulae}, taking into account that 
\eq{defq} implies that $\partial_\tau\big|_{\tau_0} = \ii\pi\, q\partial_q$. 
For instance, the first two derivatives are
\begin{equation}
\label{ders}
\begin{aligned}
\left.\frac{\partial(1/J)}{\partial\tau}\right|_{\tau_0} & = 
\frac{\ii\pi}{J_0}\,\frac{2E_6}{E_4}~,\\
\left.\frac{\partial^2(1/J)}{\partial\tau^2}\right|_{\tau_0} & = 
\frac{(\ii\pi)^2}{J_0}\,\frac{2(8E_6^2 + E_6 E_4 E_2 - 3 E_4^3)}{3E_4^2}~.
\end{aligned}
\end{equation}

On the other hand, we can compute the difference $\delta(1/J)$ 
via \eq{jtoG}, by expanding the coefficients $G_2$ and $G_3$ given in \eq{g2g3} 
with respect to the values they assume in the 
massless case, writing
\begin{equation}
\label{Gtog}
G_2 = g_2\, (2 a^2)^2\left(1 + \eta_2\right)~,~~~
G_3 = g_3\,(2 a^2)^3 \left(1 + \eta_3\right)~.
\end{equation}
It is not particularly useful here to spell $\eta_{2}$ and $\eta_3$ in detail, but 
we just remark that they depend on $u$ and on the flavour invariants. 
With easy manipulations we then find
\begin{equation}
\label{diffJJ0}
\delta\big(\frac{1}{J}\big)
= -\frac{E_6^2}{E_4^3}\left(\frac{(1 + \eta_3)^2}{(1 + \eta_2)^3} - 1\right)~.
\end{equation}
We can now equate the two different expressions of $\delta\big(\frac{1}{J}\big)$, given in
Eq.s~(\ref{tayloriJ}) and (\ref{diffJJ0}), order by order in the large-$a$ expansion that
is obtained substituting Eq.s~(\ref{tau}) and (\ref{expu}) in the first
and in the second one, respectively.

At the first non trivial order, namely $1/a^2$, 
\eq{tayloriJ} gives, through \eq{ders},
\begin{equation}
\label{da2tau}
\delta\big(\frac{1}{J}\big)
\Big|_{\frac{1}{a^2}}= -\frac{1}{J_0}\frac{E_6}{E_4}\,
\frac{h_0}{a^2}~.
\end{equation} 
On the other hand, \eq{diffJJ0} reduces to
\begin{equation}
\label{da2}
\delta\big(\frac{1}{J}\big)
\Big|_{\frac{1}{a^2}} = \frac{E_6\big(E_6^2-E_4^3\big)}{E_4^4}\,
\frac{R}{a^2} 
 = -\frac{1}{J_0}\frac{E_6}{E_4}\, \frac{R}{a^2} ~;
\end{equation} 
this is basically the same computation that leads to \eq{delta1J}.
The two expressions (\ref{da2tau}) and (\ref{da2}) are identical, since 
we have already set $h_0=R$.

At the next order, $1/a^4$, we start getting non-trivial information.
{From} \eq{tayloriJ} we get
\begin{equation}
\label{del4t}
\delta\big(\frac{1}{J}\big)
\Big|_{\frac{1}{a^4}} = 
-\frac{1}{J_0}\left(
\frac{3E_6}{2E_4}\, h_1 - 
\frac{8 E_6^2 +E_6 E_4 E_2-3 E_4^3}{12 E_4^2}\, h_0^2
\right)\frac{1}{a^4}~.
\end{equation}
With a bit of algebra, from \eq{diffJJ0} 
we get instead
\begin{equation}
\label{del4}
\delta\big(\frac{1}{J}\big)
\Big|_{\frac{1}{a^4}} = 
-\frac{1}{J_0}\left(
\frac{3 E_4^3 - 8 E_6^2 + 3 E_6 E_4(e_1 - 4\lambda(q))}{12 E_4^2}\,R^2
-\frac{3E_6}{2E_4} \left(\theta_4^4 T_1 - \theta_2^4 T_2\right)
\right)\frac{1}{a^4}~.
\end{equation}
Comparing these two expressions, and taking into account \eq{lambdais},
we find 
\begin{equation}
\label{h1is}
h_1 = \frac{E_2}{6}\,R^2-\,\theta_4^4\,T_1+\,\theta_2^4\,T_2~.
\end{equation}

At the next-to-next order, we are able to determine the coefficient $h_2$, which appears in the 
$1/a^6$ term of \eq{tayloriJ}, since the corresponding term in the expansion
of \eq{diffJJ0} contains only quantities already determined, namely $h_0,\lambda$ 
and $q\partial_q h_1$. This pattern, which is easily implemented on a symbolic
computation program like \emph{Mathematica}, continues at all orders, and allows us to
determine recursively the coefficients $h_\ell$. The explicit results up to 
$h_4$ are
\begin{align}
h_2&=\frac{1}{90}\,\Big(E_4+5E_2^2\Big)R^3+\frac{2}{5}\,E_4\,N-
\frac{1}{3}\,\theta_4^4\Big(2E_2+2\theta_2^4+\theta_4^4\Big)\,RT_1\notag\\
&~~~+\frac{1}{3}\,\theta_2^4\Big(2E_2-\theta_2^4-2\theta_4^4\Big)\,RT_2~,
\label{h2ex}\\
\phantom{\Bigg\{}h_3&=\frac{1}{7560}
\Big(11E_6+84E_4E_2+175E_2^3\Big)R^4+\frac{2}{35}\Big(3E_6+7E_4E_2
 \Big)RN\notag\\
 &~~~-\frac{1}{12}\theta_4^4\Big(3E_4+5E_2^2+8E_2\theta_2^4+4E_2\theta_4^4\Big)R^2T_1\notag\\
&~~~+\frac{1}{12}\theta_2^4\Big(3E_4+5E_2^2-4E_2\theta_2^4-8E_2\theta_4^4\Big)R^2T_2\notag\\
 &~~~-\frac{1}{14}\Big(4E_6-7E_2\theta_4^8-14\,\theta_4^{12}-28\,\theta_2^4\theta_4^8\Big)
T_1^2\notag\\
 &~~~-\frac{1}{14}\Big(4E_6-7E_2\theta_2^8+14\,\theta_2^{12}+28\,\theta_2^8\theta_4^4\Big)
T_2^2\notag\\
&~~~-\frac{1}{7}\Big(2E_6+7E_2\theta_2^4\theta_4^4+7\,\theta_2^8\theta_4^{4}-
 7\,\theta_2^4\theta_4^8\Big)T_1T_2~, 
\label{h3ex}
\\
\phantom{\Bigg\{}h_4&=\frac{1}{22680}\Big(44E_6E_2+19E_4^2+196E_4E_2^2+245E_2^4\Big)R^5\notag\\
&~~~+\frac{2}{315}\Big(36E_6E_2+20E_4^2+49E_4E_2^2
 \Big)R^2N\notag\\
 &~~~+\frac{1}{135}\theta_4^4\Big(12E_6-47E_4E_2-35E_2^3-\big(35E_2^2+30\theta_4^8\big)
\big(2\theta_2^4+\theta_4^4\big)\Big)R^3T_1\notag\\
&~~~-\frac{1}{135}\theta_2^4\Big(12E_6-47E_4E_2-35E_2^3+\big(35E_2^2+30\theta_2^8\big)
\big(\theta_2^4+2\theta_4^4\big)\Big)R^3T_2\notag\\
 &~~~+\frac{4}{15}\theta_4^4\Big(2E_6-2E_4E_2-5\theta_4^8\big(2\theta_2^4+\theta_4^4\big)\Big)
NT_1\notag\\
 &~~~-\frac{4}{15}\theta_2^4\Big(2E_6-2E_4E_2+5\theta_2^8\big(\theta_2^4+2\theta_4^4\big)\Big)
NT_2\notag
\\
&~~~-\frac{1}{63}\Big(E_6\big(24E_2-8\theta_4^4\big)-\theta_4^8\big(133E_4+49E_2^2\big)
-112E_2\theta_4^8\big(2\theta_2^4+\theta_4^4\big)\notag\\
&~~~~~~~~~~~~~~~+75\theta_4^{16}+20\theta_4^{12}\theta_2^4+4
\theta_2^{16}\Big)RT_1^2\notag\\
&~~~-\frac{1}{63}\Big(E_6\big(24E_2+8\theta_2^4\big)-\theta_2^8\big(133E_4+49E_2^2\big)
+112E_2\theta_2^8\big(\theta_2^4+2\theta_4^4\big)\notag\\
&~~~~~~~~~~~~~~~+75\theta_2^{16}+20\theta_2^{12}\theta_4^4+4
\theta_4^{16}\Big)RT_2^2\notag\\
&~~~+\frac{2}{189}\Big(76E_6E_2-147E_2^2\big(E_4-\theta_2^8-\theta_4^8\big)
+ 112 E_2\big(\theta_2^{12}-\theta_4^{12}\big)\notag\\
&~~~~~~~~~~~~~~~- 6\theta_2^{16} - 
33 \theta_2^{12}\theta_4^4 + 150\theta_2^8 \theta_4^8 - 33 \theta_2^4 
\theta_4^{12} - 6\theta_4^{16}\Big)RT_1T_2~.
\label{h4ex}
\end{align}
{From} the modular transformation properties of the Eisenstein series and the $\theta$-functions,
we see that the coefficients $h_\ell$ are almost modular forms of degree $2\ell$; the failure to
be exact modular forms is due to the appearance of the Eisenstein series $E_2$ whose modular transformations are anomalous.
Note that the recursive method we have described fixes completely the
coefficients $h_\ell$, differently from the recursion relation based on the modular
anomaly equation \cite{Minahan:1997if} which only determines the $E_2$ dependence.
In models with many different structures like the $N_f=4$ theory, this is a big 
computational advantage.
In the next section we are going to analyze these results and comment on their properties.

\section{Discussion of the results and comments}
\label{sec:results}
The explicit formulas for the coefficients $h_\ell$ 
obtained in the previous section allow us to read, via 
\eq{htochir}, the exact expression for 
the elements of the SO(8) chiral ring of the eight-dimensional theory on the D7 branes.
For example, from \eq{h1is} we have
\begin{equation}
 \vev{\,\Tr m^{4}\,}= E_2\,R^2-\,6\theta_4^4\,T_1+\,6\theta_2^4\,T_2~.
\label{ff4}
\end{equation}
Expanding the modular functions in powers of $q$, we can obtain the various instanton contributions. Explicitly,
we have
\begin{equation}
\begin{aligned}
 \vev{\,\Tr m^{4}\,}=&~\big(1 - 24 q^2 -72 q^4\big)R^2-
\big(6-48q+144q^2-192q^3+144q^4-288q^5\big)T_1\phantom{\Big|}\\
&~+\big(96q+384q^3+576q^5\big)T_2+\cO(q^6)\phantom{\Big|}\\
=&~\Tr\vev{\,m\,}^{4}- 48\, \Pf m ~q -24\sum_{i<j} m_i^2 m_j^2 ~ q^2 -192\, \Pf m~ q^3\phantom{\Bigg|}\\
&~-\Big(12\sum_im_i^4+48\sum_{i<j}m_i^2m_j^2
\Big)q^4-288\,\mathrm{Pf}m\,q^5+\cO(q^6)~,\phantom{\Big|}
\end{aligned}
\label{m4exp}
\end{equation}
where in the final step we used the definitions (\ref{invariants}) of the mass invariants.
One can check that this result completely agrees with the one obtained via localization techniques
in Ref.s~\cite{Fucito:2009rs,Billo:2009di} from direct multi-instanton calculations performed in 
the D7/D(-1) brane system of type I$^\prime$. Furthermore, the non-perturbative part 
of the quartic correlator (\ref{ff4}) matches precisely against the exact results 
for the BPS-saturated quartic coupling in the dual Heterotic string (see, for instance, the discussion in
Ref.~\cite{Billo:2009di}).

This analysis can be extended to the higher elements of the SO(8) chiral ring, and again we
find perfect agreement with all results existing in the literature on this matter. The details
for the chiral correlators up to $\vev{\,\Tr m^{10}\,}$ are given in Appendix \ref{app:chirinst},
together with their expansions up to the first few instantons.

We observe that by setting $T_1=T_2=N=0$ and retaining only the dependence on the quadratic invariant
$R$, our results reduce to those found in Ref.~\cite{Minahan:1997if} with a different method; 
indeed with these positions the SU(2)
$N_f=4$ theory reduces to the so-called $\cN=2^*$ model (also known as mass deformed $\cN=4$ SU(2) theory) that was studied in that reference.
In Appendix \ref{app:n2star} we present an alternative derivation
of the recursion relation for the $\cN=2^*$ model and also discuss the decoupling limits to the pure 
SU(2) theory.

Another interesting remark is that the coefficients $h_\ell$ given in Eq.s~(\ref{h1is}) - (\ref{h4ex})
satisfy 
\begin{equation}
\frac{\partial h_\ell}{\partial E_2} = \frac{\ell}{6}\,\sum_{m=1}^{\ell}h_{m-1}\,
h_{\ell-m}~~~~\mbox{for}~\ell\geq1~,
 \label{recursionhl}
\end{equation}
with the initial condition ${\partial h_0}/{\partial E_2}=0$. This recursion relation fixes the
$E_2$ dependence of all coefficients $h_\ell$ and could be used to reconstruct them in analogy
to what has been done for the $\cN=2^*$ SU(2) model in Ref.~\cite{Minahan:1997if}.
 
\eq{recursionhl} can be given a nice interpretation in terms of the modular anomaly equation
\cite{Minahan:1997ct,Bershadsky:1993cx}. Let us consider 
the combination $(a_D-\tau_0\,a)$, which, using Eq.s~(\ref{preptoad}) and
(\ref{expF}), can be written as
\begin{equation}
\label{ad0a}
 a_D-\tau_0\,a = \frac{1}{2\pi\ii}\sum_{\ell=0}^\infty\frac{1}{2^\ell}\,\frac{h_\ell}{a^{2\ell+1}}~,
\end{equation}
and study its modular transformation properties. We recall that under an
$S$ modular transformation we have
\begin{equation}
 \tau_0\,\to\,-\frac{1}{\tau_0}~,~~~~a_D\,\to\,-a~,~~~~
a\,\to\,a_D=\tau_0\,a\Big(1+\frac{1}{2\pi\ii\tau_0}
\sum_{\ell=0}^\infty\frac{1}{2^\ell}\,\frac{h_\ell}{a^{2\ell+2}}\Big)~;
\label{stransf}
\end{equation}
moreover we assume that the coefficients $h_\ell$ are almost modular forms of weight $2\ell$
transforming under $S$ as follows
\begin{equation}
h_\ell\,\to\,h_\ell'=\tau_0^{2\ell}\big(h_\ell+\delta h_\ell\big)~,
\label{hltransf}
\end{equation}
and that the ``anomalous'' term $\delta h_\ell$ arises only through the fact that
$h_\ell$ depends on $E_2$, {\it{i.e.}} it is of the form
\begin{equation}
\label{halphaE}
h_\ell = \alpha_0\, E_2^\ell + \alpha_{1}\, E_2^{\ell-1} + \ldots + \alpha_{\ell-1}\, E_2 + \alpha_\ell~,
\end{equation}
with $\alpha_{\ell}$ modular forms of weight $2\ell$. Taking into account the modular properties of $E_2$
(see \eq{E2mod}), this implies that
\begin{equation}
\delta h_\ell= \frac{6}{\ii\pi\tau_0} \frac{\partial h_\ell}{\partial E_2} + \cO(\tau_0^{-2})~.
\label{deltahil} 
\end{equation}
Applying these rules to \eq{ad0a}, on the one hand we have
\begin{equation}
\label{aad2}
a_D- \tau_0 \,a \to -a +\frac{1}{\tau_0}\,a_D =
\frac{1}{2\pi\ii\,\tau_0}\sum_{\ell=0}^\infty\frac{h_\ell}{2^\ell\,a^{2\ell+1}}~,
\end{equation}
while on the other hand we have
\begin{equation}
\label{aad1}
a_D- \tau_0\, a \,\to\,\frac{1}{2\pi\ii}\sum_{\ell=0}^\infty \frac{h'_\ell}{2^\ell (\tau_0\,a)^{2\ell+1}}\,
\Big(1+\frac{1}{2\pi\ii\,\tau_0}\sum_{m=0}^\infty\frac{h_m}{2^m a^{2m+2}}\Big)^{-2\ell-1}~.
\end{equation}
Comparing the right hand sides of these equations, and using the form given in 
\eq{deltahil} for $h'_\ell$, we find
\begin{equation}
\label{aad3}
\sum_{\ell=0}^\infty \frac{6}{2^\ell} \, \frac{\partial h_\ell}{\partial E_2}\,\frac{1}{a^{2\ell}}
= \sum_{m,n=0}^\infty \frac{(2m+1)}{2^{m+n+1}}\,\frac{h_m h_n}{a^{2(m+n+1)}}
\end{equation}
from which, after a suitable relabeling of the indices, the recursion relation (\ref{recursionhl}) and
its initial condition easily follow. It is interesting to notice that \eq{aad3} 
is nothing but the mode expansion of the following partial differential equation
\begin{equation}
\label{diffa}
\frac{\partial}{\partial E_2}(a_D- \tau_0\,a) = 
\frac{\pi}{6\ii}(a_D- \tau_0 \,a) \frac{\partial}{\partial a}
(a_D- \tau_0\, a)~,
\end{equation}
which is a type of inviscid Burgers' equation.

We conclude by observing that the recursive methods we have described in this paper could be generalized 
in several ways. 
In particular it would be very intersting to apply them
to models with gauge groups of higher rank corresponding to SW curves of higher genus and use them to
find further connections with their F-theory interpretation. Another interesting possibility would be
to apply these techniques to the gravitational corrections of the 4$d$ Yang-Mills theories and establish
a connection with the topological amplitudes at higher genus. Finally, it would be nice to study the
relation between the elements of the SO(8) chiral ring we have found and the corresponding amplitudes
in the dual Heterotic string, generalizing the connection already established for the first
element $\vev{\,\Tr m^{4}\,}$ in Ref.s~\cite{Billo:2009di,Fucito:2009rs} using the D-instanton
interpretation. In particular, it would be interesting to see how the modular anomaly we have found in our
expressions for the chiral ring elements might be related to the holomorphic anomaly for the dual
Heterotic amplitudes. We hope to return to some of these issues in the near future.

\vskip 1cm
\noindent {\large {\bf Acknowledgments}}
\vskip 0.2cm
We thank Francesco Fucito, Francisco Morales and Igor Pesando for very useful discussions.

\vskip 1cm
\appendix

\section{Useful formulae}
\label{app:formulae}

\paragraph{Modular functions:}
All the functions we are going to discuss depend on a modulus $\tau$ and admit a Fourier
expansion in terms of $q = \exp(\ii\pi\tau)$. To keep the formulae short, 
we do not indicate this dependence explicitly, except when some confusion is possible, 
and we write $\theta_a$ for $\theta_a(0|\tau)$, $E_2$ for $E_2(\tau)$ and so on.

The Jacobi $\theta$-functions are defined as
\begin{equation}
\label{thetadef}
 \theta\spin{a}{b}(v|\tau)=\sum_{n\in \Z} q^{(n-\frac{a}{2})^2}
\, \ee^{2\pi \ii (n-\frac{a}{2})(v-\frac{b}{2})}~,
\end{equation}
for $a,b=0,1$. We simplify the notation by writing, as usual, 
$\theta_1\equiv\theta\spin{1}{1}$,
$\theta_2\equiv\theta\spin{1}{0}$, $\theta_3\equiv\theta\spin{0}{0}$, 
$\theta_4\equiv\theta\spin{0}{1}$.
The functions $\theta_a$, $a=2,3,4$, satisfy the ``\emph{aequatio identica satis 
abstrusa}''
\begin{equation}
\label{abstruse}
\theta_3^4 - \theta_2^4 -\theta_4^4=0~.
\end{equation}

The Dedekind $\eta$-function is defined by
\begin{equation}
\label{dede}
\eta(q)= q^\frac{1}{12} \prod_{n=1}^\infty (1-q^{2n})~.
\end{equation}

The first Eisenstein series can be expressed as follows:
\begin{equation}
\label{E246}
\begin{aligned}
E_2 & = 1 - 24 \sum_{n=1}^\infty \sigma_{1}(n)\, q^{2n} 
= 1 - 24 q^2 -72 q^4 - 96 q^6 +\ldots~,\\
E_4 & = 1 + 240 \sum_{n=1}^\infty \sigma_{3}(n)\, q^{2n}
= 1 + 240 q^2 + 2160 q^4 + 6720 q^6 + \ldots~,\\
E_6 & = 1 - 504 \sum_{n=1}^\infty \sigma_{5}(n)\, q^{2n}
= 1 - 504 q^2 -16632 q^4 - 122976 q^6 + \ldots~,
\end{aligned}
\end{equation}
where $\sigma_{k}(n)$ is the sum of the $k$-th power of the divisors 
of $n$, i.e., $\sigma_k(n) = \sum_{d|n} d^k$.
The series $E_4$ and $E_6$ are expressible as polynomials in the $\theta$-functions
according to
\begin{equation}
\label{eistotheta}
\begin{aligned}
E_4 & = \frac 12 \big(\theta_2^8 + \theta_3^8 + \theta_4^8\big)~,\\
E_6 & = \frac 12 \big(\theta_3^4 + \theta_4^4\big)
 \big(\theta_2^4 + \theta_3^4\big) \big(\theta_4^4 - \theta_2^4\big)~. 
\end{aligned}
\end{equation}
The series $E_2$, $E_4$ and $E_6$
are connected among themselves by logarithmic $q$-derivatives
and form a sort of a ``ring'': 
\begin{equation}
\label{Eisring}
\begin{aligned}
q\partial_q E_2 & = \frac 16\big(E_2^2-E_4\big)~,\\
q\partial_q E_4 & = \frac{2}{3}\big(E_4 E_2-E_6\big)~,\\
q\partial_q E_6 & = E_6 E_2-E_4^2~.
\end{aligned}
\end{equation}
Also the derivatives of the functions $\theta_a^4$ have simple expressions:
\begin{equation}
\label{dertheta}
\begin{aligned}
q\partial_q \theta_2^4 & = \frac{\theta_2^4}{3}
\big(E_2+\theta_3^4 + \theta_4^4\big)~,\\
q\partial_q \theta_3^4 & = \frac{\theta_3^4}{3}
\big(E_2+\theta_2^4 -\theta_4^4\big)~,\\
q\partial_q \theta_4^4 & = \frac{\theta_4^4}{3}
\big(E_2-\theta_2^4 -\theta_3^4 \big)~.
\end{aligned}
\end{equation}

\paragraph{Modular transformations:}
Under a Sl$(2,\mathbb Z)$ modular transformation
\begin{equation}
\label{modular}
\tau\to \tau' = \frac{a\tau + b}{c\tau+d}~,
\end{equation}
the Eisenstein series $E_{4}$ and $E_6$ are modular forms
of weight $4$ and $6$, respectively:
\begin{equation}
\label{modforms}
E_4(\tau') = (c\tau + d)^4 \,E_4(\tau)~,~~~
E_6(\tau') = (c\tau + d)^6 \,E_6(\tau)~.
\end{equation}
The series $E_2$, instead, is an almost modular form of degree 2:
\begin{equation}
\label{E2mod}
E_2(\tau') = (c\tau + d)^2 \,E_2(\tau) - \frac{6\ii}{\pi}\,c\,(c\tau + d)~. 
\end{equation}
The behaviour of the relevant $\theta$-functions and the Dedekind function 
under the generators $T$ and $S$ of the modular group is given by
\begin{equation}
\begin{aligned}
&T~:~~~~ \theta_3^4 \leftrightarrow \theta_4^4 ~,~~~ 
\theta_2^4 \to \theta_2~,~~~
\eta\to \ee^{\frac{\ii\pi}{12}} \,\eta~,\phantom{\Bigg|}\\
&S~:~~~~ \theta_2^4  \to \tau^2\, \theta_4^4~,~~~
\theta_3^4  \to \tau^2 \,\theta_3^4~,~~~
 \theta_4^4  \to \tau^2 \,\theta_2^4~,~~~
\eta\to \sqrt{-\ii\tau}\, \eta~.
\end{aligned}
\label{modtran1}
\end{equation}

\section{Chiral ring elements and their instanton expansion}
\label{app:chirinst}
Here we write the exact expressions for the first few elements of the SO(8) chiral ring (beyond 
$\vev{\,\Tr m^{4}\,}$ that we already discussed in Section \ref{sec:results}), together 
with their expansions up to the first few instantons. The formulas we are going to write
follow from Eq.s~(\ref{h2ex}) - (\ref{h4ex})
and \eq{htochir}), as well as the definitions (\ref{invariants}) of the
mass invariants. We have
 \begin{align}
\vev{\,\Tr m^{6}\,}&=\frac{1}{12}\,\Big(E_4+5E_2^2\Big)R^3+3\,E_4\,N-
\frac{5}{2}\,\theta_4^4\Big(2E_2+2\theta_2^4+\theta_4^4\Big)\,RT_1\notag\\
&~~~+\frac{5}{2}\,\theta_2^4\Big(2E_2-\theta_2^4-2\theta_4^4\Big)\,RT_2\label{m6ex}\\
\phantom{\Bigg\{}&=\Tr\vev{\,m\,}^{6}+ 180\sum_{i<j<k} m_i^2 m_j^2
 m_k^2 ~q^2 + 960\, \Pf m \sum_i m_i^2~ q^3 \notag\\
&~~~+\Big(180\sum_{i\not= j}m_i^4 m_j^2 + 2160\!\sum_{i<j<k}m_i^2 m_j^2
 m_k^2\Big)\,q^4+5760\,\Pf m\sum_i m_i^2~ q^5+\ldots~,\notag
\\
\phantom{\Bigg\{}\vev{\,\Tr m^{8}\,}  &=\frac{1}{1080}
\Big(11E_6+84E_4E_2+175E_2^3\Big)R^4+\frac{2}{5}\Big(3E_6+7E_4E_2
 \Big)RN\notag\\
 &~~~-\frac{7}{12}\theta_4^4\Big(3E_4+5E_2^2+8E_2\theta_2^4+4E_2\theta_4^4\Big)R^2T_1\notag\\
&~~~+\frac{7}{12}\theta_2^4\Big(3E_4+5E_2^2-4E_2\theta_2^4-8E_2\theta_4^4\Big)R^2T_2\notag\\
 &~~~-\frac{1}{2}\Big(4E_6-7E_2\theta_4^8-14\,\theta_4^{12}-28\,\theta_2^4\theta_4^8\Big)
T_1^2\notag\\
 &~~~-\frac{1}{2}\Big(4E_6-7E_2\theta_2^8+14\,\theta_2^{12}+28\,\theta_2^8\theta_4^4\Big)
T_2^2\notag\\
&~~~-\Big(2E_6+7E_2\theta_2^4\theta_4^4+7\,\theta_2^8\theta_4^{4}-
 7\,\theta_2^4\theta_4^8\Big)T_1T_2
\label{m8ex}
\\
\phantom{\Bigg\{}&=\Tr\vev{\,m\,}^{8}- 420\, (\Pf m)^2 ~q^2 - 2240\,\Pf m\,
 \sum_{i<j}m_i^2 m_j^2~ q^3 \notag\\
&~~~-\Big(210\sum_{i<j} m_i^4 m_j^4 + 4200 \sum_i m_i^4\!\!\sum_{j<k\not=
 i}m_j^2 m_k^2 + 40320 \,(\Pf m)^2\Big)\,q^4\notag\\
&~~~- \Pf m\Big(9408\sum_i m_i^4 + 53760 \sum_{i<j} m_i^2 m_j^2\Big)\,q^5+\ldots~, 
\notag\\
\phantom{\Bigg\{}\vev{\,\Tr m^{10}\,}&=\frac{1}{4032}\Big(44E_6E_2+19E_4^2+196E_4E_2^2+245E_2^4\Big)R^5\nonumber\\
&~~~+\frac{1}{28}\Big(36E_6E_2+20E_4^2+49E_4E_2^2
 \Big)R^2N\nonumber\\
 &~~~+\frac{1}{24}\theta_4^4\Big(12E_6-47E_4E_2-35E_2^3-\big(35E_2^2+30\theta_4^8\big)
\big(2\theta_2^4+\theta_4^4\big)\Big)R^3T_1\nonumber\\
&~~~-\frac{1}{24}\theta_2^4\Big(12E_6-47E_4E_2-35E_2^3+\big(35E_2^2+30\theta_2^8\big)
\big(\theta_2^4+2\theta_4^4\big)\Big)R^3T_2\nonumber\\
 &~~~+\frac{3}{2}\theta_4^4\Big(2E_6-2E_4E_2-5\theta_4^8\big(2\theta_2^4+\theta_4^4\big)\Big)
NT_1\nonumber\\
 &~~~-\frac{3}{2}\theta_2^4\Big(2E_6-2E_4E_2+5\theta_2^8\big(\theta_2^4+2\theta_4^4\big)\Big)
NT_2\nonumber
\\
&~~~-\frac{5}{56}\Big(E_6\big(24E_2-8\theta_4^4\big)-\theta_4^8\big(133E_4+49E_2^2\big)
-112E_2\theta_4^8\big(2\theta_2^4+\theta_4^4\big)\nonumber\\
&~~~~~~~~~~~~~~~+75\theta_4^{16}+20\theta_4^{12}\theta_2^4+4
\theta_2^{16}\Big)RT_1^2\nonumber\\
&~~~-\frac{5}{56}\Big(E_6\big(24E_2+8\theta_2^4\big)-\theta_2^8\big(133E_4+49E_2^2\big)
+112E_2\theta_2^8\big(\theta_2^4+2\theta_4^4\big)\nonumber\\
&~~~~~~~~~~~~~~~+75\theta_2^{16}+20\theta_2^{12}\theta_4^4+4
\theta_4^{16}\Big)RT_2^2\nonumber\\
&~~~+\frac{5}{84}\Big(76E_6E_2-147E_2^2\big(E_4-\theta_2^8-\theta_4^8\big)
+ 112 E_2\big(\theta_2^{12}-\theta_4^{12}\big)\nonumber\\
&~~~~~~~~~~~~~~~- 6\theta_2^{16} - 
33 \theta_2^{12}\theta_4^4 + 150\theta_2^8 \theta_4^8 - 33 \theta_2^4 
\theta_4^{12} - 6\theta_4^{16}\Big)RT_1T_2 
\label{m10ex}
\\
\phantom{\Bigg\{}&=\Tr\vev{\,m\,}^{10}+3360\,\Pf m\!\sum_{i<j<k} m_i^2 m_j^2 m_k^2~q^3
\notag\\
&~~~+\Big(5670\sum_i m_i^2 \sum_{j<k\not= i}m_j^4 m_k^4 + 52920\, (\Pf
 m)^2 \sum_i m_i^2 \Big)\,q^4\notag\\
&~~~+\Pf m\,\Big(60480\sum_{i \not= j} m_i^4 m_j^2 + 362880\sum_{i<j<k} m_i^2
 m_j^2 m_k^2\Big)\,q^5+\ldots~.
\end{align}
These results agree with those found with explicit multi-instanton
calculations in Ref.s~\cite{Billo:2009di,Fucito:2009rs,Billo:2010mg}. 

\section{Recursion relation for the $\cN=2^*$ theory}
\label{app:n2star}
The SW curve for the $\cN=2^*$ theory with gauge group $SU(2)$
corresponds to a particular case of the $N_f=4$ theory where the 
invariants $T_i$ and $N$ vanish. In this case \eq{curvemassive} gets the simple factorized form  
\begin{equation}
 y^2= (x - W_1)(x-W_2)(x-W_3)=
 \big(x-e_1\tilde{u}-e_1^2 R\big)\big(x-e_2\tilde{u}-e_2^2 R\big)\big(x-e_3\tilde{u}-e_3^2 R\big)~,
\end{equation}
so that one can establish a recursion relation based on \eq{kappais}.
Indeed the anharmonic ratio of the roots is found to be
\begin{equation}
\label{anharm2*}
\kappa = \frac{W_3 - W_2}{W_1 - W_2} = \kappa_0 
\left(1 - \frac{\theta_4^4\,R}{u-\theta_2^4\, R/2}\right)~,
\end{equation}  
where
\begin{equation}
\label{kappa0}
\kappa_0 = \frac{e_3 - e_2}{e_1 - e_2} = \frac{\theta_2^4}{\theta_3^4}
\end{equation}
is the corresponding ratio of the roots in the massless case.

On the one hand, we can use \eq{anharm2*} to express the difference 
$(\kappa - \kappa_0)$ in terms of $u$, and then plug in the large-$a$ expansion
of the latter in the form of \eq{expu}. 
On the other hand, we can Taylor expand $\kappa$, seen as the
function of $\tau$ given by \eq{kappais}, around $\kappa_0$:
 \begin{equation}
\label{taylorkappa}
\kappa - \kappa_0 = \sum_{n=1}^\infty \frac{1}{n!} 
\left.\frac{\partial^n \kappa}{\partial\tau^n}\right|_{\tau_0} (\tau - \tau_0)^n~ ,
\end{equation} 
and then insert the large-$a$ expansion of $(\tau - \tau_0)$ given in \eq{tau}.

Comparing order by order the two different expansions of $(\kappa-\kappa_0)$ we 
can recursively determine the unknown coefficients $h_\ell$ of the expansion 
exactly in $q$ and in $R$, finding in the end that  
\begin{eqnarray}
 \phantom{\Bigg|}&&h_0=R~,~~~ h_1 = \frac{E_2}{6}\,R^2~,~~~ h_2 = \frac{E_4 + 5 E_2^2}{90}\, R^3~,~~~
 h_3 = \frac{11 E_6 + 84 E_4 E_2 + 175 E_2^3}{7560}\, R^4~,\notag\\
 \phantom{\Bigg|}&&h_4=\frac{44E_6E_2+19E_4^2+196E_4E_2^2+245E_2^4}{22680}\,R^5~,\ldots~.
 \label{hfromn2*}
 \end{eqnarray}
 These expressions correspond to setting $T_i=N=0$ in the $N_f=4$ results given in 
 Eq.s~(\ref{h1is}) - (\ref{h4ex}), and coincide with what was found in Ref.~\cite{Minahan:1997if}.

\paragraph{Decoupling limits to the pure SU(2) theory:}
As already discussed in Ref.~\cite{Seiberg:1994aj}, it is possible to recover the
pure $\mathrm{SU}(2)$ theory from the $\cN=2^*$ model
by sending the mass invariant $R$ to infinity and at the same time $q$ to 
zero, so as to keep the combination
\begin{equation}
\label{n2*topure}
\hat\Lambda^2 = 2 R q 
\end{equation}
finite.  Indeed, expanding the $\theta$-functions, in this limit from \eq{anharm2*} we find
\begin{equation}
\label{kappared}
\kappa \to - \frac{16 q R}{u - 8 q R} =
\frac{-2\hat\Lambda^2}{\hat u-\hat\Lambda^2}
\end{equation}
which coincides with the result (\ref{kappasu2}) derived from the SW curve (\ref{puresu2})
of the pure SU(2) theory. Notice that above we
have taken into account the fact that $u = 4 \hat u$ according to what we have explained after
\eq{tauYM2}. 
 
Let us now consider the complex structure $\tau$ given in \eq{tau}; using the $h_\ell$'s
of \eq{hfromn2*}, it is not difficult to check that in this limit one
gets
\begin{equation}
\label{limtauhatpure}
\tau\to \hat\tau = \frac{\ii}{\pi} \log \frac{4a^2}{{\hat \Lambda}^2} +
\frac{1}{2\pi\ii} \left\{\frac{3}{2} \frac{{\hat\Lambda}^4}{a^4}
+ \frac{105}{64} \frac{{\hat\Lambda}^8}{a^8}
+  \ldots \right\}~,
\end{equation} 
which is the correct expression of the effective $\mathrm{SU}(2)$ coupling in the 
normalization (\ref{tauYM2}) appropriate for the form \eq{kappasu2} of the Matone relation.

Starting from the generic $N_f=4$ theory one can decouple some masses and
recover the asymptotically free theories with $N_f=3,2,1,0$. In particular, 
one can reach the pure SU(2) case by sending $q\to 0$ while keeping
\begin{equation}
\label{pflim}
\Lambda^4 =  32\, q\,\Pf m  
\end{equation}
fixed. Such a limit can be taken starting from a particular form of 
the $N_f=4$ curve (\ref{curvemassive}) in which $R=N=T_1=0$ and
only $T_2 = -\Pf m/2$ is kept; in this situation, the curve
factorizes and the anharmonic ratio of the roots becomes
\begin{equation}
 \label{anhR}
\kappa = \frac{(\theta_2^4 + \theta_3^4)u - \theta_4^4\sqrt{u^2  - 2 
\theta_2^4\theta_3^4 \Pf m}}{(\theta_2^4 + \theta_3^4)u + \theta_4^4\sqrt{u^2  -
2 \theta_2^4\theta_3^4 \Pf m}}~.
\end{equation}
In the decoupling limit one gets indeed
\begin{equation}
 \label{kpflim}
\kappa \to \frac{u - \sqrt{u^2 - 32 q\Pf m}}{u + \sqrt{u^2 - 32 q\Pf m}}
= \frac{u - \sqrt{u^2 - \Lambda^4}}{u + \sqrt{u^2 - \Lambda^4}}~,
\end{equation}
which agrees with the pure SU(2) result mentioned in footnote \ref{foot_norm}. 
In this case, for the complex structure
$\tau$ we find
\begin{equation}
 \label{taulimpf}
\tau \to \frac{2\ii}{\pi} \log \frac{8\sqrt{2}a^2}{\Lambda^2} + \frac{1}{\pi\ii} 
\left\{\frac{3}{16} \frac{\Lambda^4}{a^4}
+ \frac{105}{4096} \frac{\Lambda^8}{a^8}
+  \ldots\right\}~,
\end{equation}
which, with the position $\hat\Lambda^4 = \Lambda^4/8$, corresponds to
\emph{twice} the effective coupling  $\hat\tau$ of \eq{limtauhatpure}, as
appropriate for this case.

\providecommand{\href}[2]{#2}\begingroup\raggedright\endgroup

\end{document}